\documentclass[twocolumn,aps]{revtex4}
\usepackage[latin9]{inputenc}
\usepackage{amsmath}
\usepackage{amssymb}
\usepackage{graphicx}

\makeatletter

\providecommand{\tabularnewline}{\\}

\@ifundefined{textcolor}{}
{%
 \definecolor{BLACK}{gray}{0}
 \definecolor{WHITE}{gray}{1}
 \definecolor{RED}{rgb}{1,0,0}
 \definecolor{GREEN}{rgb}{0,1,0}
 \definecolor{BLUE}{rgb}{0,0,1}
 \definecolor{CYAN}{cmyk}{1,0,0,0}
 \definecolor{MAGENTA}{cmyk}{0,1,0,0}
 \definecolor{YELLOW}{cmyk}{0,0,1,0}
}

\usepackage{color}

\def\NOT(#1,#2){\OneQubitGate(#1,#2){$X$}}

\makeatother

\begin{document}

\title{ Improved indirect control of nuclear spins in diamond NV centers}

\author{Jingfu Zhang, Swathi S. Hegde and Dieter Suter\\
 Fakultaet Physik, Technische Universitaet Dortmund,\\
 D-44221 Dortmund, Germany }

\date{\today}
\begin{abstract}
Hybrid quantum registers consisting of different types of qubits offer
a range of advantages as well as challenges. The main challenge is
that some types of qubits react only slowly to external control fields,
thus considerably slowing down the information processing operations. 
One promising approach that has been tested in a number of cases is to use indirect control,
where external fields are applied only to qubits that interact strongly with resonant excitation pulses.
Here we use this approach to indirectly control the nuclear spins of an NV center, using microwave pulses to drive
the electron spin, combined with free precession periods optimized for generating logical gate operations on the nuclear spins. 
The scheme provides universal control and we present two typical applications:
polarizing the nuclear spin and measuring nuclear spin free induction
decay signals, both without applying radio-frequency pulses. This
scheme is versatile as it can be implemented over a wide range of
magnetic field strengths and at any temperature. 
\end{abstract}
\maketitle

\section{Introduction}

Hybrid quantum systems \cite{kurizki2015quantum}, such as electron-nuclear
spins of the nitrogen vacancy (NV) center in diamond \cite{Suter201750},
have emerged as useful physical systems for implementing quantum computing
and imaging \cite{nielsen,Stolze:2008xy,ladd2010quantum,blencowe2010quantum,cai2014hybrid}.
The difference in the inherent properties of the two subsystems are
often useful, e.g. for implementing fast gate operations on the electron
spin and achieving long information storage in the nuclear spin. However,
it also provides challenges for the coherent control of hybrid spin
systems, e.g. since the interaction between the nuclear spin magnetic
moment and the control fields is orders of magnitude weaker than that
of the electron spins, which has relatively short coherence times.

The magnetic moment of the nuclear spin can be enhanced by the hyperfine
coupling with the electron spin in the NV center \cite{Suter201750,PhysRevB.94.060101}.
This corresponds to an enhancement in the effective gyromagnetic ratio
of the nuclear spin and is determined by the strength of the coupling.
For systems with sufficiently strong couplings, like those of the
nearest $^{13}$C spins to the NV center \cite{PhysRevB.94.060101,PhysRevA.87.012301,PhysRevLett.102.210502},
or the $^{14}$N spin in the NV center \cite{dobrovitski2012,cnotN15,chen15},
the enhancement results in 
 nuclear spin Rabi frequencies 
 that are high enough for direct control of the nuclear spins 
by the application of radio-frequency fields. However, the enhancement
and thus the usefulness of the nuclear spins as qubits decreases dramatically
with increasing distance of the nuclear spin from the NV center.

 To alleviate this problem, the  approach relying on
indirect control of the nuclear spins was developed theoretically \cite{PhysRevA.76.032326,PhysRevA.91.042340} and experimentally demonstrated in various  systems, 
like the malonic acid radical \cite{PhysRevA.78.010303,PhysRevLett.107.170503} and NV centers \cite{PhysRevLett.109.137602,naturephoton,PhysRevB.96.134314,Pan13}. 
This scheme does not require any radio-frequency pulses and therefore
achieves much faster operations on the nuclear spins. Instead, it
uses only microwave (MW) pulses acting on the electron spin, combined
with free precession under the effect of anisotropic hyperfine interactions.
The anisotropic interactions 
 result in different orientations of the nuclear spin quantization
axes for different states of the electron spin, and provide the possibility
to achieve indirect control of the nuclear spin by only controlling
the electron spin \cite{PhysRevA.76.032326,PhysRevA.78.010303}. The
basic idea of this scheme is to start with the electron spin initialized,
e.g., in the $m_{S}=0$ state and the nuclear spin aligned along the
corresponding quantization axis, which is close to the $z$-axis.
If a MW pulse changes the state of the electron spin on a timescale
that is fast compared to the precession period of the nuclear spin,
the nuclear spin remains unchanged during the MW pulse. After the
pulse, it is therefore oriented at a nonzero angle from the new quantization
axis and starts to precess. The control procedure then consists in
finding the best combination of precession periods around the different
axes that bring the spin close to the targeted orientation. The control
efficiency can be improved by maximizing the angle between the 
different quantization axes of the nuclear spin \cite{PhysRevA.76.032326}.

In our present work, we chose a system where the hyperfine couplings
are close to the Larmor frequency of the nuclear spin, which optimizes
the difference in the orientation of the quantization axes to a value
close to $90^{\circ}$. Moreover, unlike the scheme in \cite{PhysRevA.76.032326},
we generalized the switched control scheme by replacing the 
$180^{\circ}$ pulses by arbitrary operations implemented through
pulses with variable flip angles and phases. 
Such generalization extends the search space for optimizing the pulse
sequence parameters, and is useful to improve the control efficiency,
e.g., to reduce the number of pulses. In our work, the elementary
unitary operations consist of only 2 - 3 rectangular MW pulses separated
by delays. Compared to earlier works based on dynamical decoupling
\cite{naturephoton,PhysRevB.96.134314} or modulated microwave pulses
\cite{PhysRevA.78.010303,PhysRevLett.107.170503} that used hundreds
or even thousands of MW pulse segments, this is a dramatic reduction
of the control cost.

In this article, we use this scheme to implement operations that occur
in many quantum information or imaging tasks: we generate and detect
nuclear spin coherence, transfer population between the electronic
and nuclear spins and generate a pseudo-Hadamard gate on the nuclear
spin. Using these operations, we polarize the nuclear spin and measure
the nuclear spin transition frequencies using free induction decay
(FID) signals. Our system of interest consists of the electron spin
and one $^{13}$C nuclear spin, which is relatively weakly coupled
to the electron ($A$\textless 0.2 MHz). In this system, the nuclear
spin transition frequencies (NMR spectrum) are spread over a spectral
range of more then 200 kHz, which would be out of reach for direct
radio-frequency excitation but can be readily excited by our indirect
control scheme.

\section{System and operations to implement}

The spin system consists of the electron, the $^{14}$N nuclear spin
and one $^{13}$C nuclear spin. In this context, we do not consider
the $^{14}$N spin but focus on the subsystem where the $^{14}$N
is (and remains) in the $m_{N}$=1 state. The spins interact with
a weak magnetic field $B$ oriented along the symmetry axis of the
NV-center. In a suitable reference frame (for details see Appendix
A), we can write the relevant part of the Hamiltonian as 
\[
\frac{\mathcal{H}_{e,C}}{2\pi}=DS_{z}^{2}-(\nu_{e}-A_{N})S_{z}-\nu_{C}I_{z}+A_{zz}S_{z}I_{z}+A_{zx}S_{z}I_{x},
\]
where $S_{z}$ denotes the electron spin-1 operator, and $I_{x/z}$
the $^{13}$C spin-1/2 operators. The zero-field splitting is $D=2.87$
GHz and $\nu_{e/C}=\gamma_{e/C}B$ denote the Larmor frequencies of
the electron and $^{13}$C nuclear spins and $\gamma$ the gyromagnetic
ratios. $A_{N}=-2.16$ MHz is the secular part of the hyperfine coupling
with the $^{14}$N nuclear spin while $A_{zz}$ and $A_{zx}$ are
the relevant components of the $^{13}$C hyperfine tensor.

The eigenstates of $\mathcal{H}_{e,C}$ are $|1,\varphi_{+}\rangle,|1,\psi_{+}\rangle,|0,\uparrow\rangle,|0,\downarrow\rangle,|-1,\varphi_{-}\rangle$,
and $|-1,\psi_{-}\rangle$, where $|\uparrow\rangle$ and $|\downarrow\rangle$
are the eigenstates of $I_{z}$, 
\begin{eqnarray}
|\varphi_{\pm}\rangle & = & |\uparrow\rangle\cos(\theta_{\pm}/2)+|\downarrow\rangle\sin(\theta_{\pm}/2)\nonumber \\
|\psi_{\pm}\rangle & = & -|\uparrow\rangle\sin(\theta_{\pm}/2)+|\downarrow\rangle\cos(\theta_{\pm}/2)\label{eigenC}
\end{eqnarray}
are the nuclear-spin eigenstates and
\begin{equation}
\theta_{\pm}=\arctan\frac{A_{zx}}{A_{zz}\mp\nu_{C}}\label{eq2}
\end{equation}
are the angles between the nuclear spin quantization axis and the
$z$-axis of our coordinate system for the subsystems where the electron
spin is in the state $m_{S}=\pm1$. $|1\rangle$, $|0\rangle$ and
$|-1\rangle$ are the eigenstates of $S_{z}$. Additional details
are given in Appendix A. The nuclear spin transition frequencies are
$\nu_{C}$ and $\nu_{\pm}=\sqrt{A_{zx}^{2}+(\nu_{C}\mp A_{zz})^{2}}$
if the electron spin is in the state $m_{S}=0$, and $m_{S}=\pm1$,
respectively. In the following, we use the eigenstates $\{|0\rangle,|-1\rangle\}\otimes\{|\uparrow\rangle,|\downarrow\rangle\}$
of the operators $S_{z}$ and $I_{z}$ as our computational basis.

The experiments were performed at room temperature. We used a $^{12}$C-enriched
diamond crystal with a $^{13}$C concentration of $0.002\%$ \cite{PhysRevLett.110.240501,1882-0786-6-5-055601}
and applied a magnetic field $B=14.8$ mT. We selected a center with
a resolved coupling to a $^{13}$C nuclear spin, with the coupling
constants $A_{zz}=-0.152$ MHz and $A_{zx}=0.110$ MHz. For this center,
the quantization axis of the nuclear spin is oriented at an angle
$\theta_{+}=-10^{{\circ}}$, $\theta_{-}=86^{{\circ}}$ and $\theta_{0}\approx0$
from the $z$-axis if the electron spin is in the $\pm1$ or 0 state.
The MW pulses had a Rabi frequency of $\approx0.5$ MHz, which is
small compared to $A_{N}$ and to $\nu_{e}$. Accordingly, they only
drive the transition from the $m_{S}=0$ to one of the $m_{S}=\pm1$
states and are selective for $m_{N}=1$.

\begin{figure}[t]
\includegraphics[width=1\columnwidth]{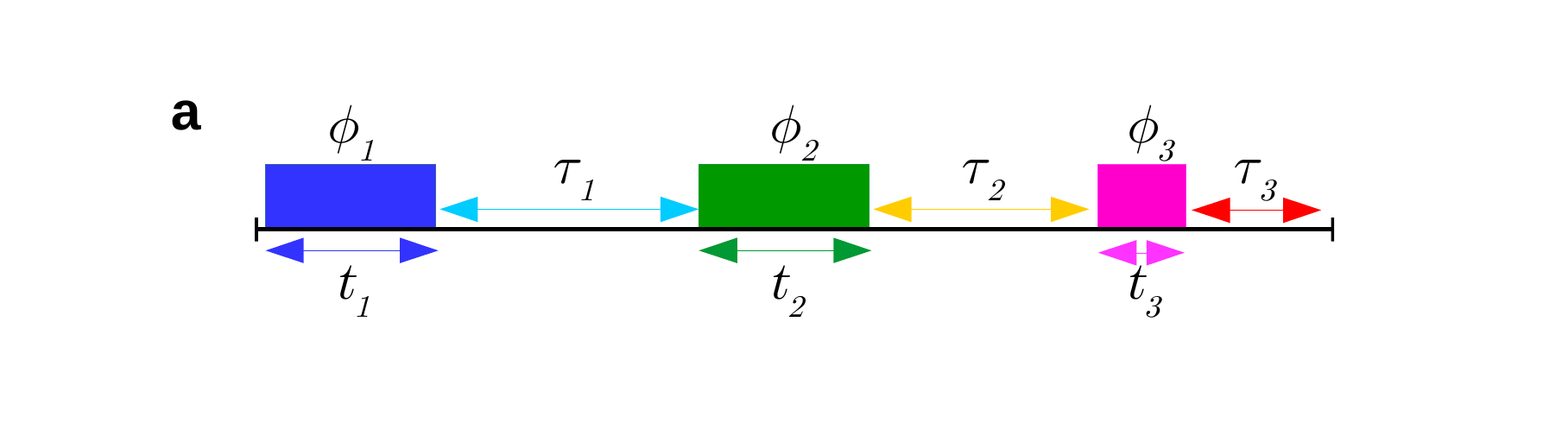} \\
 \includegraphics[width=1\columnwidth]{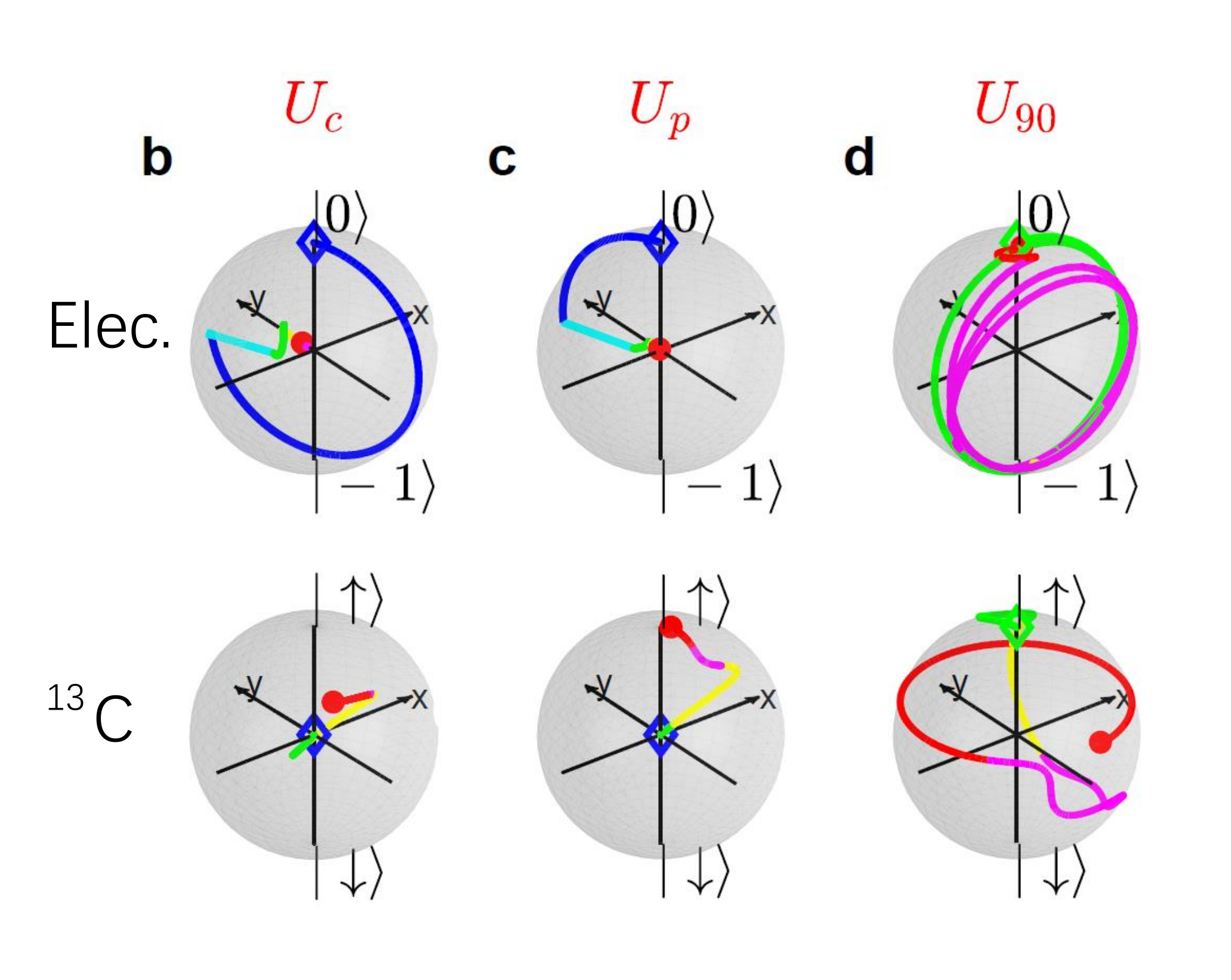} 
 \caption{Pulse sequence and Bloch sphere representation for the evolution of
the electron- and $^{13}$C spins. (a) MW pulse sequence at a fixed
power with the carrier frequency set to the ESR transition frequency
$m_{S}=0\leftrightarrow-1$. The pulse durations, phases and delays
are indicated as $t_{k}$, $\phi_{k}$, and $\tau_{k}$ respectively.
(b-d) Evolution trajectories of the electron and $^{13}$C spin on
the Bloch sphere under the operations $U_{c}$, $U_{p}$ and $U_{90}$.
The initial and final states are indicated by diamonds and filled
circles, respectively. In Fig. (b), $U_{c}$ maps $\rho_{0}$ to $\rho_{c}$,
corresponding to a mixed electron state and a maximum possible $^{13}$C
coherence. In (c), $U_{p}$ swaps the states between the electron
and $^{13}$C. In (d), the electron and $^{13}$C spins are both initially
in state $|0\rangle$. Under $U_{90}$, the $^{13}$C spin evolves
towards the -$y$-axis, while the electron returns to its initial
state. Here $U_{90}$ consists of only 2 MW pulses and 2 delays, i.e.,
$t_{1}$ =$\tau_{1}$ = 0 in Fig. (a). The color of each partial trajectory
corresponds to the color in (a). The parameters of these operations
are given in Appendix B. The $x$, $y$ and $z$ components of the
final states on the Bloch sphere are presented in Table \ref{blochxyz}
in Appendix D.}
\label{bloch}
\end{figure}

As initial examples of the indirect control approach, we implement
three state-to-state transfer operations to which we refer as $U_{c}$,
$U_{c}^{\dagger}$ , and $U_{p}$, and one unitary operation, $U_{90}$.
The initial state for the first three operations is
\begin{equation}
\rho_{0}=|0\rangle\langle0|\otimes E/2,\label{inistate}
\end{equation}
where $E$ is the $2\times2$ identity operator. This state corresponds
to the electron spin in a pure state and the $^{13}$C nuclear spin
in the maximally mixed state.

We first consider the operations $U_{c}$ and $U_{c}^{\dagger}$,
where $U_{c}$ converts $\rho_{0}$ into nuclear spin coherence in
the manifolds $m_{S}=0$ and $m_{S}=-1$ of the electron spin and
$U_{c}^{\dagger}$ transfers the coherence back to population in $m_{S}=0$.
The target state of the operation $U_{c}$ is
\begin{equation}
\rho_{c}=U_{c}\rho_{0}U_{c}^{\dagger}=\frac{1}{2}(|0\rangle\langle0|\otimes|s_{0}\rangle\langle s_{0}|+|-1\rangle\langle-1|\otimes|s_{-}\rangle\langle s_{-}|)\label{rohTp}
\end{equation}
where
\begin{eqnarray}
|s_{0}\rangle & = & (|\uparrow\rangle+i|\downarrow\rangle)/\sqrt{2}\nonumber \\
|s_{-}\rangle & = & (|\varphi_{-}\rangle-|\psi_{-}\rangle)/\sqrt{2}\label{supeigenC}
\end{eqnarray}
denote superpositions of the nuclear spin in the energy eigenbasis
of the manifolds $m_{S}=0$ and $-1$.

$U_{p}$ implements a selective population transfer. It transforms
the initial state $\rho_{0}$ to 
\begin{equation}
\rho_{p}=U_{p}\rho_{0}U_{p}^{\dagger}=\frac{1}{2}(|0\rangle\langle0|+|-1\rangle\langle-1|)\otimes|\uparrow\rangle\langle\uparrow|,\label{roh2}
\end{equation}
where the electron spin is in the maximally mixed state in the subspace
$m_{S}=\{0,-1\}$ while the nuclear spin is in the pure state $|\uparrow\rangle$.
Clearly $U_{p}$ implements a SWAP operation that aligns the nuclear
spin along the $z$-axis.

To implement the operation $U_{c}$, $U_{c}^{\dagger}$ or $U_{p}$,
we search for pulse sequence that maximizes the overlap between the
final state and the target state. We use a MATLAB$^{\tiny{\textregistered}}$
subroutine based on a genetic algorithm \cite{Mitchell:1998:IGA:522098}
to find the optimal set of parameters. The third operation $U_{90}=e^{-i(\pi/2)I_{x}}$
is a $\pi/2$ rotation around the $x$-axis for $^{13}$C, or a pseudo-Hadamard
gate, since it is equivalent to the Hadamard gate $U_{H}$ up to operations
around the $z$-axis as 
\begin{equation}
U_{H}=ie^{-i(\pi/2)I_{z}}U_{90}e^{-i(\pi/2)I_{z}},\label{shad}
\end{equation}
where the $z$- rotations can be easily absorbed in a suitable reference
frame shift \cite{PhysRevLett105200402,PRA78012328}. For $U_{90}$,
we maximize the process fidelity of the gate. Fig. \ref{bloch} (a)
illustrates the pulse sequence, where the pulse durations $t_{k}$,
phases $\phi_{k}$, and the durations $\tau_{k}$ of the free precession
periods were used as adjustable parameters in the optimization.

To obtain sequences that are robust against fluctuations of the MW
power, we averaged the fidelities over a range of MW field amplitudes,
as described in the Appendix B. We obtained theoretical state fidelities
of $95\%$ for the target states of $U_{c}$ and $U_{c}^{\dagger}$,
and $98\%$ for $U_{p}$ with sequences of 3 pulses and 3 delays,
and gate fidelity of $92\%$ for $U_{90}$ with a sequence of 2 pulses
and 2 delays. The total duration of these pulse sequences is $\approx7-15\mu$s,
shorter than the transverse relaxation time $T_{2}^{*}\approx20\mu$s
of the electron spin. Figs. \ref{bloch}(b-d) show the evolution of
the electron and $^{13}$C spins during the pulse sequence on the
Bloch sphere during these pulse sequences.

\section{Experimental results}

\begin{figure*}
\centering{}\centering{} \includegraphics[width=0.8\columnwidth]{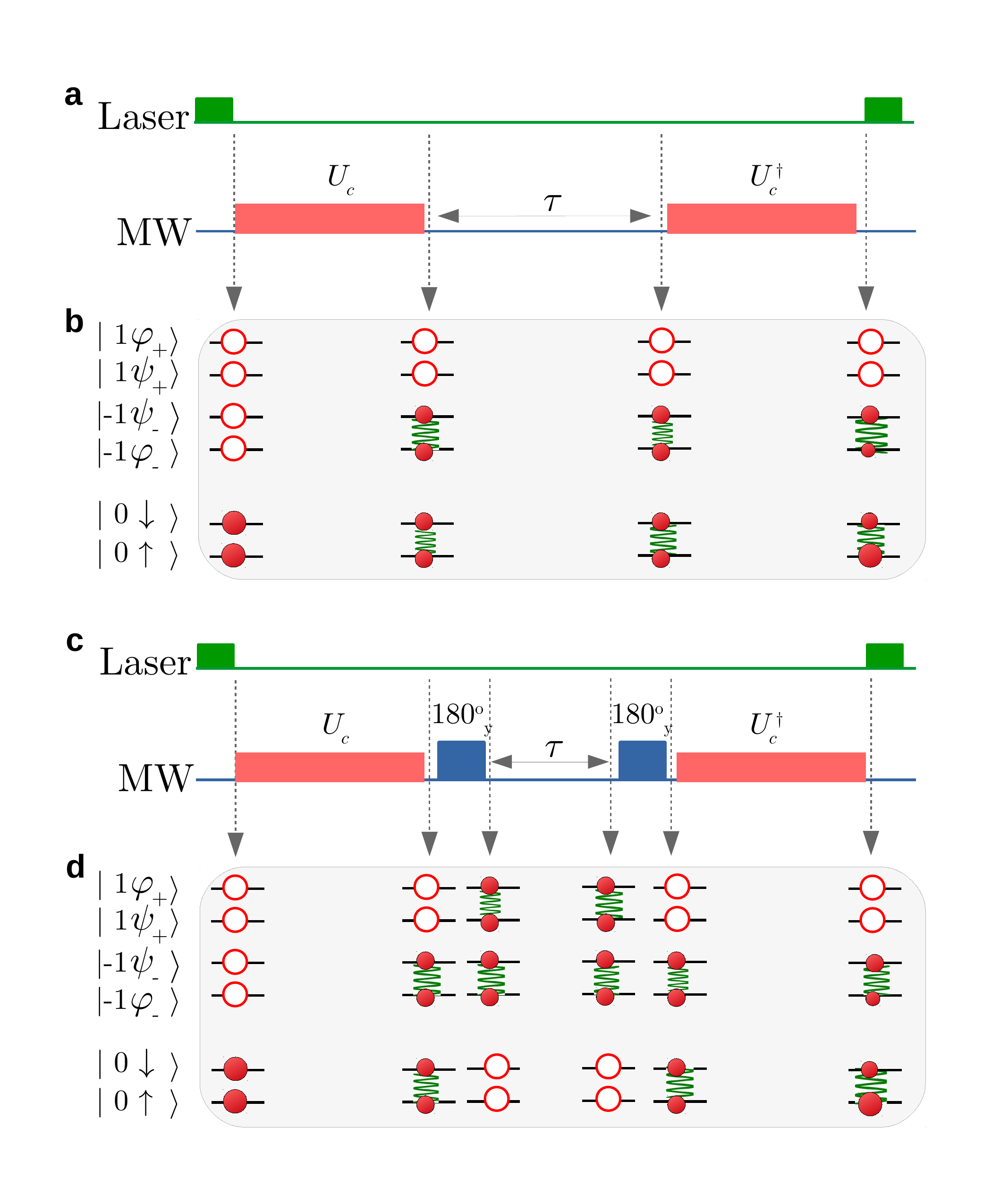}
\includegraphics[width=1.2\columnwidth]{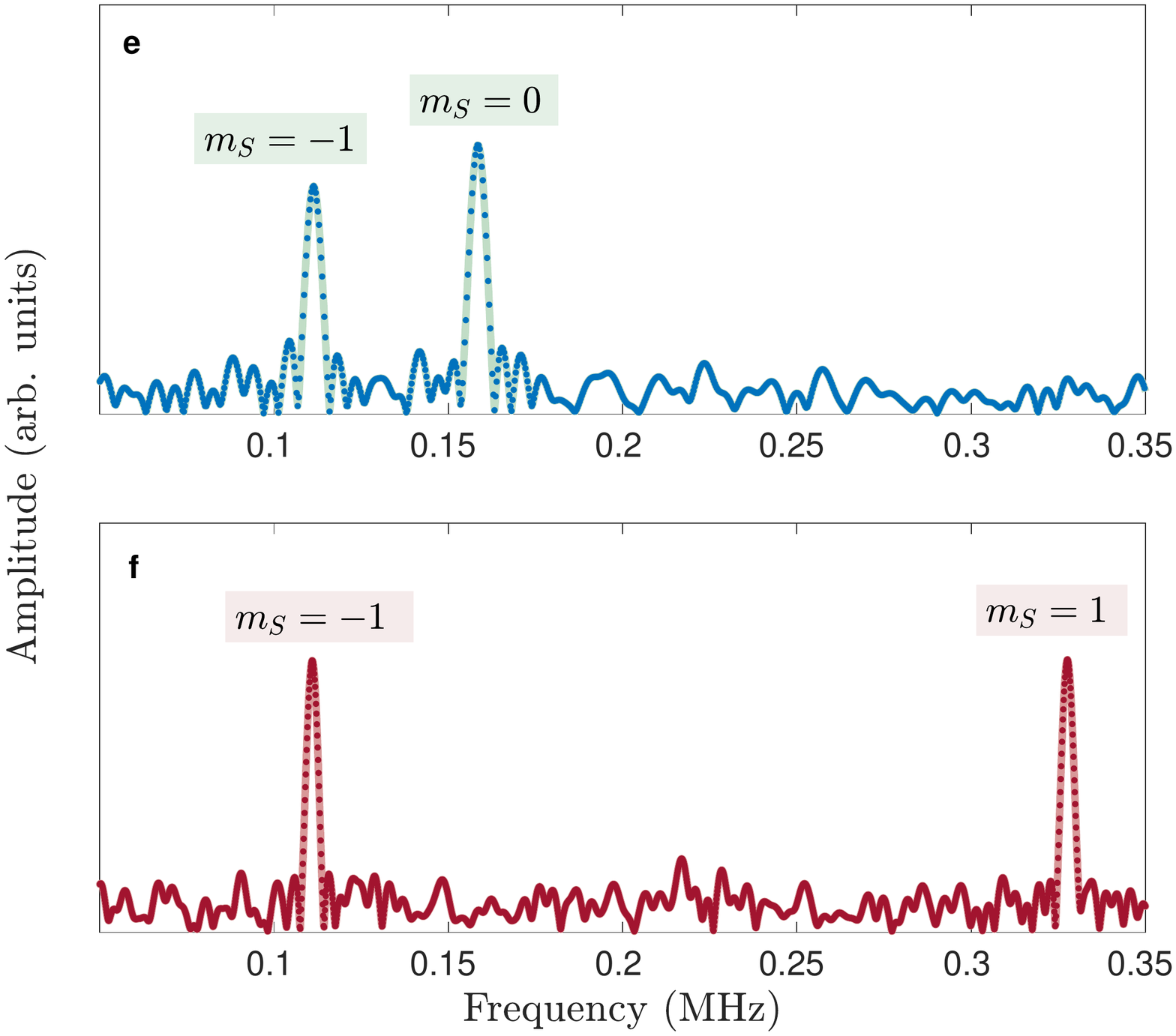} \caption{Experimental schemes and results for generating and detecting coherence
in $^{13}$C spin via indirect control. (a, c) Pulse sequences. The
red rectangles denote MW pulse sequences resonant with transition
$m_{S}=0\leftrightarrow-1$ to implement the operations $U_{c}$ and
$U_{c}^{\dagger}$, as indicated above the rectangles. The blue rectangles
denote $180^{\circ}$ -pulses resonant with transition $m_{S}=0\leftrightarrow1$
in the $m_{N}=1$ subspace. The first laser pulse initializes the
electron and $^{13}$C in state $\rho_{0}$. $U_{c}$ transforms $\rho_{0}$
to $\rho_{c}$. $\rho_{c}$ or $\rho_{c}^{\prime}$ then undergoes
free evolution for time $\tau$ and $U_{c}^{\dagger}$ transfers the
evolved coherence back to population of the $m_{S}=0$ state. During
the second laser pulse, photon counting measures this population.
(b, d) Schematic representations of the state evolution at each stage
of the operation. The filled circles represent the populations and
the wavelike patterns represent coherence. (e-f) Spectra of the $^{13}$C
spin, measured with the sequences (a) and (c), respectively. The states
of electron spin are indicated in the panels for each resonance line.}
\label{figfidsingle} 
\end{figure*}

The experimental scheme to measure $^{13}$C transition frequencies
via indirect FID measurement in electron subspaces of $m_{S}=\{0,-1\}$,
and the pictorial representations of the state evolution are shown
in Figs. \ref{figfidsingle} (a, b). The first laser pulse initializes
the system to state $\rho_{0}$. This is followed by the generation
of $^{13}$C coherence using $U_{c}$ as indicated by the wavelike
patterns in Fig. \ref{figfidsingle} (b). $\rho_{c}$ then evolves
freely for a time $\tau$. $U_{c}^{\dagger}$ converts the final coherence
back to population of $m_{S}=0$. The last laser pulse is used to
measure the population of $m_{S}=0$, and generates a signal proportional
to 
\begin{equation}
P_{|0\rangle}=Tr\{\rho_{c}\rho_{\tau}\}=[\cos(2\pi\nu_{C}\tau)+\cos(2\pi\nu_{-}\tau)]/8+1/4\label{eq:popUc}
\end{equation}
 with $\rho_{\tau}=e^{-i\tau\mathcal{H}_{e,C}}\rho_{c}e^{i\tau\mathcal{H}_{e,C}}$.

Figs. \ref{figfidsingle} (c-d) show the experimental scheme and state
representations for the $^{13}$C FID in $m_{S}=\{-1,1\}$. It starts
with the same sequence as in (a) to put the system in to the state
$\rho_{c}$. A first $180^{\circ}$ pulse applied to the $m_{S}=0\leftrightarrow1$
transition then transforms $\rho_{c}$ into 
\begin{equation}
\rho_{c}^{\prime}=(|1\rangle\langle1|\otimes|s_{0}\rangle\langle s_{0}|+|-1\rangle\langle-1|\otimes|s_{-}\rangle\langle s_{-}|)/2.\label{rohTp1}
\end{equation}

In the $m_{S}=1$ state, the nuclear spin quantization axis is almost
parallel to the $z$-axis ($\theta_{+}\approx0^{\circ}$). Therefore,
the nuclear spin coherence remains an almost equal weight superposition
of the two eigenstates, which subsequently undergoes free evolution
for a time $\tau$. After the free evolution, the second $180^{\circ}$
pulse exchanges again the states $m_{S}=1$ and $m_{S}=0$ and $U_{c}^{\dagger}$
works in the same manner as in Fig. \ref{figfidsingle} (a). The signal
generated after the last laser pulse is proportional to
\begin{equation}
P_{|0\rangle}^{\prime}=Tr\{\rho_{c}^{\prime}\rho_{\tau}^{\prime}\}=[\cos(2\pi\nu_{-}\tau)+\cos(2\pi\nu_{+}\tau)]/8+1/4\label{eq:popUcp}
\end{equation}
 with $\rho_{\tau}^{\prime}=e^{-i\tau\mathcal{H}_{e,C}}\rho_{c}^{\prime}e^{i\tau\mathcal{H}_{e,C}}$.

Figs. \ref{figfidsingle} (e-f) show the resulting $^{13}$C spectra,
obtained by Fourier transformation of the FID data which are presented
in Fig. \ref{fid_expa4} in the Appendix D. Since $\rho_{c}$ and
$\rho_{c}^{\prime}$ contain coherence in two different NMR transitions,
each of the resulting spectra features two resonance lines. The measured
transition frequencies are 0.159, 0.111, and 0.328 MHz and agree well
with the analytical solutions for $\nu_{C}$, $\nu_{-}$ and $\nu_{+}$
respectively. The linewidths are not the natural linewidths but are
determined by the truncation of the FID signal. Since the FID for
Fig. \ref{figfidsingle} (f) was measured for 300 $\mu$s and that
for (e) for 200 $\mu$s, the resonance lines in (f) are slightly narrower.
More details are presented in Appendix D.

\begin{figure}[t]
\centering{}\includegraphics[width=1\columnwidth]{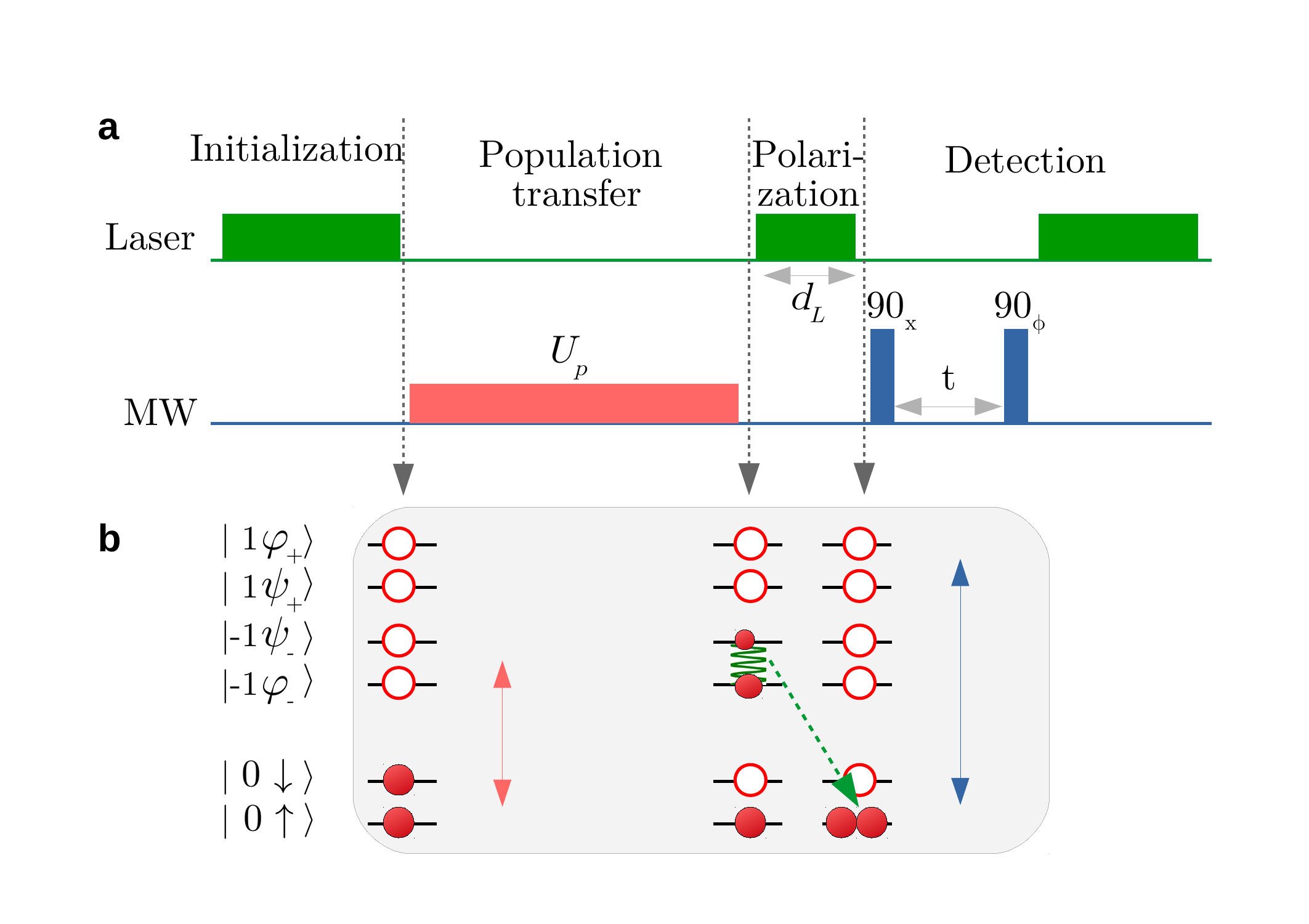}\\
 \includegraphics[width=0.85\columnwidth]{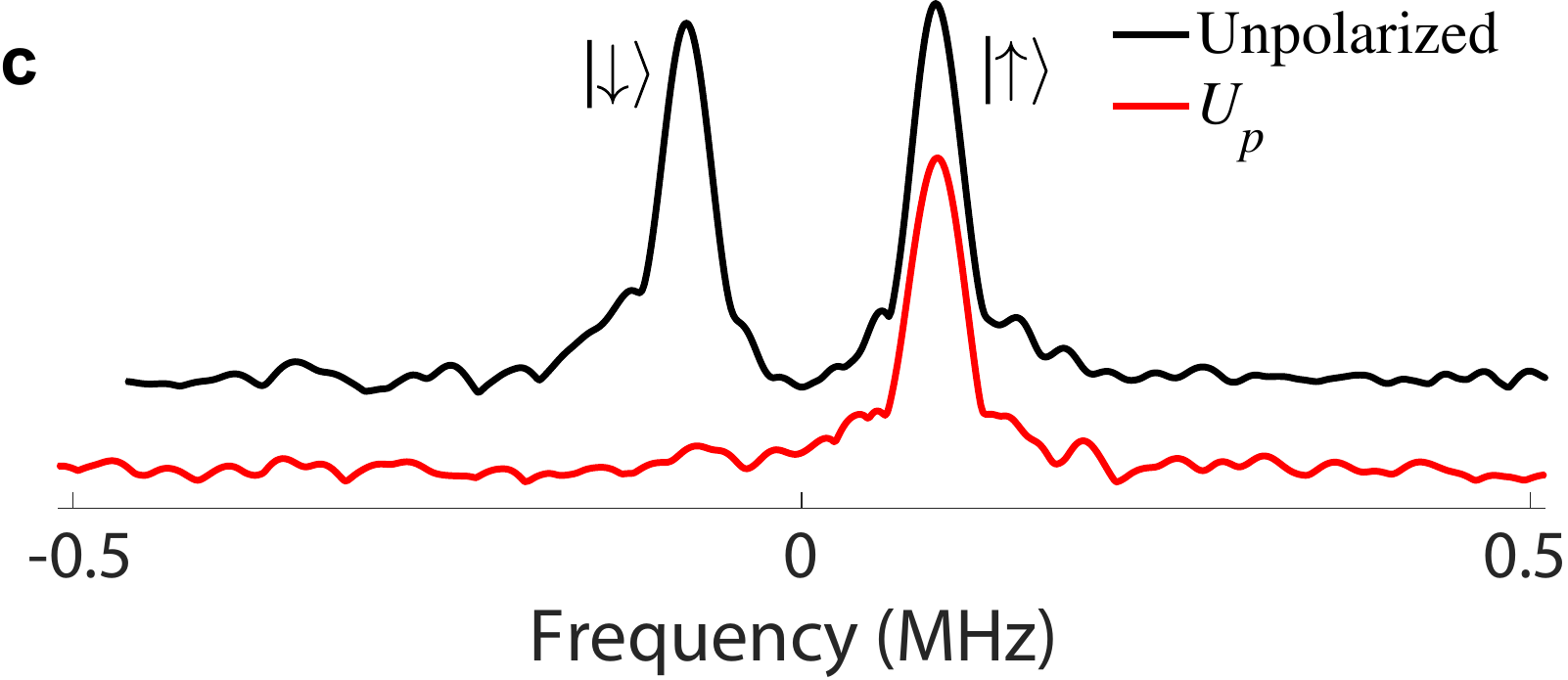} \caption{Polarization scheme and experimental results. (a) Pulse sequence.
In the initialization step, the electron and $^{13}$C spins are set
into state $\rho_{0}$. For the population transfer, we apply MW pulses
to transfer the population from states $|0,\downarrow\rangle$ to
$|-1,\uparrow\rangle$. In the polarization step, a laser pulse resets
the electron from $m_{S}=-1$ back to $m_{S}=0$. We then measure
an electron spin FID on the transition $m_{S}=0\leftrightarrow m_{S}=1$
to determine the populations of the carbon spin states $|\uparrow\rangle$
and $|\downarrow\rangle$. Here we use MW pulses with a Rabi frequency
of 3.7 MHz to reduce the operation time, since the selectivity for
the subspace of $^{14}$N state is not required. The peaks of other
$^{14}$N states are also observed but not shown here. (b) Schematic
representation of the state during the intermediate steps in (a).
The filled circles denote the population size. The short (long) vertical
double arrows indicate the transition $m_{S}=0\leftrightarrow-1$
($m_{S}=0\leftrightarrow1$). (c) ESR spectra of the $m_{S}=0\leftrightarrow1$
transition obtained from states $\rho_{0}$ and $\rho_{p}$ when $d_{L}$
=0, where the $^{13}$C spin is unpolarized (thermal) and in state$|\uparrow\rangle$,
indicated by the black and red curves, respectively.}
\label{figpulseOC} 
\end{figure}

\begin{figure}
\includegraphics[width=1\columnwidth]{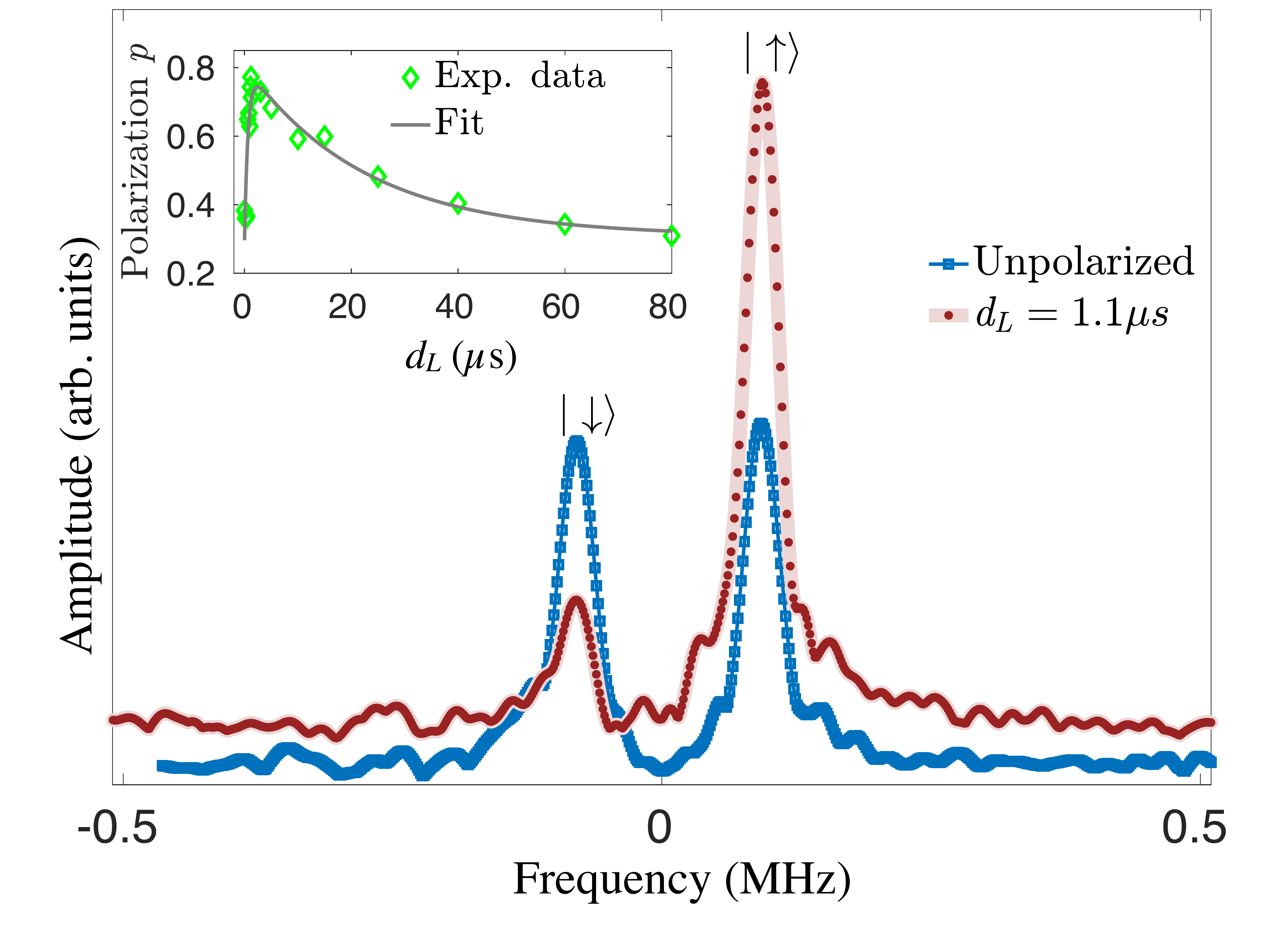} \caption{ESR spectra of the $m_{S}=0\leftrightarrow1$ transition when the
$^{13}$C spin is in the unpolarized state and when it is polarized
with a laser pulse of duration $d_{L}=1.1$ $\mu$s resulting in a
polarization $p\approx80\%$. In inset shows the dependence of the
population difference $p$ on the laser pulse duration $d_{L}$.}
\label{figrepol} 
\end{figure}

As yet another illustration, we use $U_{p}$ to polarize $^{13}$C
- a necessary step in realizing quantum computation \cite{Cramer:2016aa,PhysRevLett.110.060502,Appl.Phys.Lett.105.242402,PhysRevA.87.012301,NJP19073030,arXiv:1806.05881,naturephoton,PhysRevLett.109.137602}.
Figs. \ref{figpulseOC} (a-b) show the experimental scheme and a schematic
representation of the states during the intermediate steps. After
the initialization step, $U_{p}$ transforms $\rho_{0}$ to $\rho_{p}$,
see Eq. (\ref{roh2}). The laser pulse in the polarization step resets
the electron from $m_{S}=-1$ back to $m_{S}=0$. To measure the populations
of states $|0,\uparrow\rangle$ and $|0,\downarrow\rangle$, denoted
as $P_{|0\uparrow\rangle}$ and $P_{|0\downarrow\rangle}$, in the
$m_{S}=0$ subsystem, we performed the standard FID experiment on
the electron spin, followed by a detection laser pulse. In the obtained
ESR spectrum, the amplitudes of the resonance lines are proportional
to the populations $P_{|0\uparrow\rangle}$ and $P_{|0\downarrow\rangle}$,
respectively.

Fig. \ref{figpulseOC} (c) shows the spectra obtained from $\rho_{0}$
(as a reference) and $\rho_{p}$ ,when the laser pulse in the polarization
step was switched off. In the spectrum from $\rho_{p}$, the population
of state $|0,\downarrow\rangle$ was almost completely moved away,
shown as the negligibly small left peak. Ideally, the right peak should
have the same height as in the unpolarized state spectrum. The ratio
of the two peak amplitudes is $\approx$0.92, which we use as a measure
of the fidelity of the implemented $U_{p}$.

Fig. \ref{figrepol} shows the experiment result for polarizing the
$^{13}$C spin with a laser pulse of duration $d_{L}=1.1$ $\mu$s.
The corresponding spectrum for the unpolarized $^{13}$C spin is also
shown as a reference. The inset shows the nuclear spin polarization
$p=P_{|0\uparrow\rangle}-P_{|0\downarrow\rangle}$ as a function of
the laser pulse duration, for a laser power of about 0.5 mW. It can
be fitted by the function 
\begin{equation}
p_{repo}=0.31-0.51e^{-(\alpha+\beta)d_{L}}+0.50e^{-2\gamma d_{L}}\label{poparfit}
\end{equation}
where $\alpha=1.10$ $\mu$s$^{-1}$, $\beta=0.41$ $\mu$s$^{-1}$
and $\gamma=0.022$ $\mu$s$^{-1}$ are the pumping rates for states
$|-1,\uparrow\rangle\rightarrow|0,\uparrow\rangle$ (or $|-1,\downarrow\rangle\rightarrow|0,\downarrow\rangle$),
$|-1,\uparrow\rangle\rightarrow|0,\downarrow\rangle$ (or $|-1,\downarrow\rangle\rightarrow|0,\uparrow\rangle$),
and $|0,\uparrow\rangle\leftrightarrow|0,\downarrow\rangle$, respectively
\cite{PhysRevA.87.012301}. The highest polarization of $p_{max}\approx80\%$
was reached for a laser pulse duration $d_{L}$ around $1.1\mu$s.

\begin{figure*}
\centering %
\begin{tabular}{c}
\includegraphics[width=1.4\columnwidth]{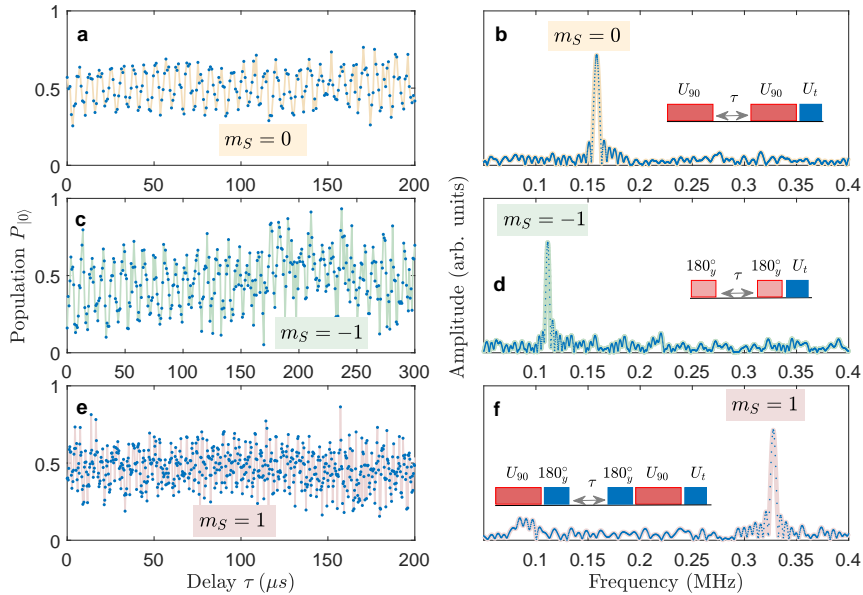} \tabularnewline
\end{tabular}\caption{Experimental $^{13}$C FIDs obtained with the transformation $U_{90}$.
The time domain data and the Fourier transforms are shown as the left
to right columns. From top to bottom we list the results obtained
in the subspaces $m_{S}=0,-1,1$, and the insets show the MW pulse
sequences correspondingly. $U_{t}$ performs as a controlled NOT -like
gate approximately in the subspace spanned in the basis $\{m_{S}=0,m_{S}=1\}\otimes\{\uparrow,\downarrow\}$
to transfer the population from state $|0,\downarrow\rangle$ to $|1,\downarrow\rangle$\cite{Ournewpaper},
so that the $^{13}$C coherence can be detected by the readout laser.
$U_{t}$ is implemented by two $90^{\circ}$ MW pulses with $90^{\circ}$
phase shift and separated by a delay of $1/(2|A_{zz}|)$.}
\label{fid_exp} 
\end{figure*}

In the following, we use this polarized state to demonstrate the pseudo-Hadamard
gate $U_{90}$. We detect its effect by implementing the standard
$^{13}$C FID experiment (see, e.g., \cite{PhysRevA.87.012301,dobrovitski2012}).
The MW pulse sequences and experimental results are shown in Fig.
\ref{fid_exp}. Unlike in subspace $m_{S}=0$, we replace $U_{90}$
by a $180_{y}^{\circ}$ MW pulse that transforms $m_{S}=0$ to $-1$,
and generates a coherence between $|\psi_{-}\rangle$ and $|\varphi_{-}\rangle$
of $^{13}$C in $m_{S}=-1$ subspace since $\theta_{-}\approx90^{\circ}$.
The scheme for the $m_{S}=1$ subspace is similar to the case of $m_{S}=0$,
except that we transfer the spin states between $m_{S}=0$ and $m_{S}=1$
by$180_{y}^{\circ}$ pulses before and after the free evolution time
$\tau$. The measured transition frequencies are measured as $0.158$,
$0.110$ and $0.328$ MHz, matching well with $\nu_{C}$, $\nu_{-}$
and $\nu_{+}$.

The performance of the operations $U_{c}$ and $U_{90}$ can be evaluated
by analyzing the signals shown in Figs. \ref{figfidsingle} and \ref{fid_exp},
combined with numerical simulation. The experimental fidelities for
$U_{c}$ and $U_{90}$ are $0.91$ and $0.74$, respectively. The
details are presented in Appendix C. The theoretical infidelities
for the sequences are 0.02, 0.05, and 0.08, for $U_{c}$, $U_{p}$
and $U_{90}$. The excess infidelities can be attributed to relaxation
effects of the electron spin and pulse imperfections.

\section{Discussion}

\subsection{The coupled $^{14}$N}

Since our interests in the current work focuses on the control of
the electron and $^{13}$C spins, perturbing effects from the $^{14}$N,
which is also coupled to the electron spin should be minimized. For
this purpose, we have chosen a suitable strength of the MW pulses,
such that the Rabi frequency of the MW pulses is strong enough compared
with the couplings of the $^{13}$C, but weak enough to affect only
one subspace of the $^{14}$N \cite{Ournewpaper}. This strategy works
well for the case where the couplings of the $^{13}$C are even weaker,
and the MW strength can be further reduced to improve the selectivity
for the subspace \cite{TranSel}. However, for the case where the
couplings of the $^{13}$C are comparable or even higher than the
coupling from the $^{14}$N, the selection of the subspace becomes
more challenging. For this case, one of the alternatives is to polarize
the $^{14}$N \cite{PolN14Duan1,PolN14Duan2}, e.g., to the state
$m_{N}=1$, so that the $^{14}$N only contributes a fixed frequency
shift, and high power (or hard) MW pulses can be used. Obviously the
effects of the coupled $^{14}$N depend on the achieved polarization
of the $^{14}$N. Recent results show that a polarization of$>98\%$
can be reached \cite{arXiv:1806.05881}.

We used numerical simulations to investigate the fidelity and duration
of the operations $U_{p}$ and $U_{90}$ for different Rabi frequencies.
The results are listed in Table \ref{tabRabi}. For $U_{p}$, the
results are not sensitive to the Rabi frequency. For $U_{90}$, higher
Rabi frequencies lead to higher fidelities, with similar total sequence
length. In these simulations, the pulse sequences are not robust against
the fluctuation of the MW power, and therefore the theoretical fidelities
at 0.5 MHz Rabi frequency are higher than the robust - sequences used
in the experiment.

\begin{table}
\begin{tabular}{|c|c|c|c|c|}
\hline 
 & \multicolumn{1}{c}{$U_{p}$ } &  & \multicolumn{1}{c}{$U_{90}$ } & \tabularnewline
\hline 
Rabi (MHz)  & Fidelity  & Duration ($\mu$s)  & Fidelity  & Duration ($\mu$s) \tabularnewline
\hline 
0.2  & 0.996  & 8.65  & 0.82  & 13.9 \tabularnewline
\hline 
0.5  & 0.997  & 7.85  & 0.97  & 13.4 \tabularnewline
\hline 
10  & 0.998  & 7.65  & 0.99  & 13.1 \tabularnewline
\hline 
\end{tabular}\caption{Simulated results for the dependence of the fidelity and total duration
of the operations $U_{p}$ and $U_{90}$ on the Rabi frequency of
the MW pulses.}
\label{tabRabi}
\end{table}

\subsection{Comparison with the previous work}

\begin{table}
\begin{tabular}{|c|c|c|c|c|}
\hline 
 & \multicolumn{1}{c}{$U_{p}$ } &  & \multicolumn{1}{c}{$U_{90}$ } & \tabularnewline
\hline 
Rabi (MHz)  & Fidelity  & Duration ($\mu$s) & Fidelity  & Duration ($\mu$s)\tabularnewline
\hline 
10  & 0.65  & 6.5  & 0.99  & 13.1 \tabularnewline
\hline 
0.5  & 0.61  & 10-17  & 0.97  & 13.5 \tabularnewline
\hline 
\end{tabular} \caption{Results obtained by simulation using the theoretical scheme proposed
in Ref. \cite{PhysRevA.76.032326}. In the optimization for $U_{p}$
or $U_{90}$, we used the same numbers of the pulses and delays in
the pulse sequence shown in Fig. \ref{Uc_pulse}, except that we fixed
the flip angles for each pulse to $180^{\circ}$.}
\label{tabswitch} 
\end{table}

In previous works based on dynamical decoupling (DD) \cite{naturephoton,PhysRevB.96.134314},
multiple cycles of the DD sequences were applied. The delay time between
the DD pulses was determined by DD spectroscopy, and the weak coupling
condition was used, where the hyperfine coupling is small compared
to the Larmor frequency of the nuclear spin. In our system, however,
the Larmor frequency $\nu_{C}$ is close to the hyperfine coupling,
violating this condition.

In Ref. \cite{Pan13}, the proximal $^{13}$C spin was studied, where
the hyperfine coupling is much larger than the Larmor frequency. In
this case, it is possible to generate a large angle between the nuclear
quantization axes of the manifolds $m_{S}=0$ and $m_{S}=-1$ by choosing
a suitable angle between the static field and the NV axis. The DD
pulses were used to flip the states of the electron spin. Compared
with this work, the hyperfine coupling in our work is much weaker,
and the Rabi frequency of the nuclear spin cannot be observed in our
experimental setup.

Using the theoretical procedure outlined in in Ref. \cite{PhysRevA.76.032326},
we simulated the short sequences of MW pulses but fixed the flip angle
for each pulse to $180^{\circ}$. We checked the optimization for
$U_{p}$ and $U_{90}$. The results of the simulation are listed in
Table \ref{tabswitch}. For $U_{p}$, the fidelity is quite low, which
indicates that this method does not work for $U_{p}$. For $U_{90}$,
however, we obtained useful results as listed in Table \ref{tabswitch}.
These results illustrated the similarity between our pulse opitimization
and the switched control in Ref. \cite{PhysRevA.76.032326}. Such similarity
can also be noticed in the pulse sequence for $U_{90}$ used in the
experiment, shown in Fig. \ref{Uc_pulse} in Appendix B , where the
flip angles are effectively close to $180^{\circ}$.

\subsection{Number of MW pulses}

Using more pulses and delays provides additional parameters for the
optimization of the pulse sequences and can therefore result in higher
theoretical fidelities. In a typical implementation, however, they
lead to longer sequences and therefore aggravate losses of the fidelity
through decoherence. Moreover, the additional pulses also introduce
additional errors from pulse imperfections. In the present system,
we found that sequences with 2-3 pulses are a good compromise.

Using numerical simulations, we investigated the dependence of the
fidelity and operation duration on the number of MW pulses. For $U_{90}$,
using 3 pulses, we can improve the fidelity to 0.99. This result is
consistent with the theoretical prediction \cite{PhysRevA.76.032326}.
However, the operation duration is about 32 $\mu$s, twice as long
as with 2 pulses (see Fig. \ref{Uc_pulse} in Appendix B), and exceeds
the transverse relaxation time of the electron spin. The control pulses
may thus have to be combined with DD pulses, as illustrated in previous
works \cite{naturephoton,PhysRevB.96.134314,Pan13}.

The number of pulses that is required to implement a specific operation
depends on the angle between the nuclear spin quantization axes in
the different electron spin eigenstates. We demonstrate this with
a simulation for the present system: if we change the static magnetic
field to 28 mT, the $^{13}$C Larmor frequency becomes 0.3 MHz and
the angle $\theta_{-}$ changes to $36.6^{\circ}$, while $\theta_{0}$
does not change. Table \ref{Tabsmall} lists the results for the operation
$U_{p}$ for a Rabi frequency of $\omega_{1}=0.5$ MHz. Here, the
optimal number of pulses is 5, which is consistent with the theoretical
prediction that number of rotations required is not more than 6 \cite{PhysRevA.76.032326}.
The pulse sequences are shown as Fig. \ref{Figsmall} in Appendix
B.

\begin{table}
\begin{tabular}{|c|c|c|}
\hline 
Number of pulses  & Fidelity  & Duration ($\mu$s) \tabularnewline
\hline 
3  & 0.86  & 11.6 \tabularnewline
\hline 
4  & 0.976  & 13.5 \tabularnewline
\hline 
5  & 0.997  & 8.89 \tabularnewline
\hline 
\end{tabular}\caption{Results of the optimisation of $U_{p}$, for $\theta_{-}=36.6^{\circ}$. }
\label{Tabsmall} 
\end{table}

\section{Conclusion}

We have demonstrated highly efficient control of nuclear spins in
a solid-state system without using any radio-frequency irradiation.
Instead, we relied on suitably chosen sequences of microwave pulses
that drive the electronic spin and thereby modulate the anisotropic
interaction and the effective field acting on the nuclear spins. The
scheme was verified for the example of diamond NV-centers, working
at room temperature. Using this technique, we implemented several
fundamental unitary operations for quantum computing, such as generating
quantum coherence, transferring populations, and Hadamard-like gate.
For this demonstration, we only used 2 or 3 MW control pulses, resulting
in short gate times. Our scheme does not require a specific choice
of the magnetic field, it can be used at arbitrary temperature and
applied to different types of hybrid qubit systems.

\textit{Acknowledgement} This work was supported by the DFG through
grants SU 192/34-1 and SU 192/19-2. We thank Daniel Burgarth for helpful
discussions .

\section*{Appendix}

\subsection{System, Hamiltonian and ESR Spectra}

\begin{figure*}[h]
\centering 
\includegraphics[width=0.9\columnwidth]{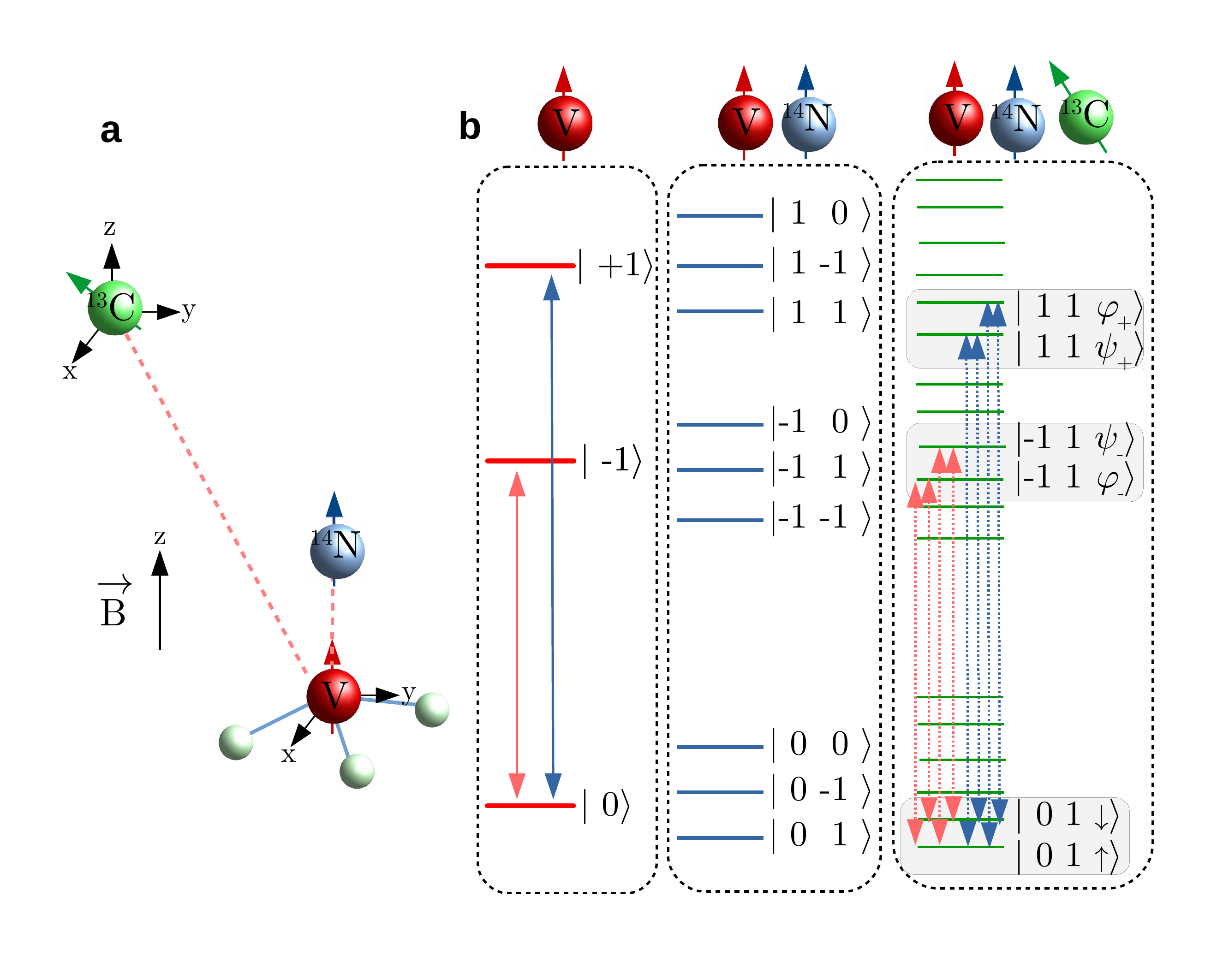} 
\includegraphics[width=1\columnwidth]{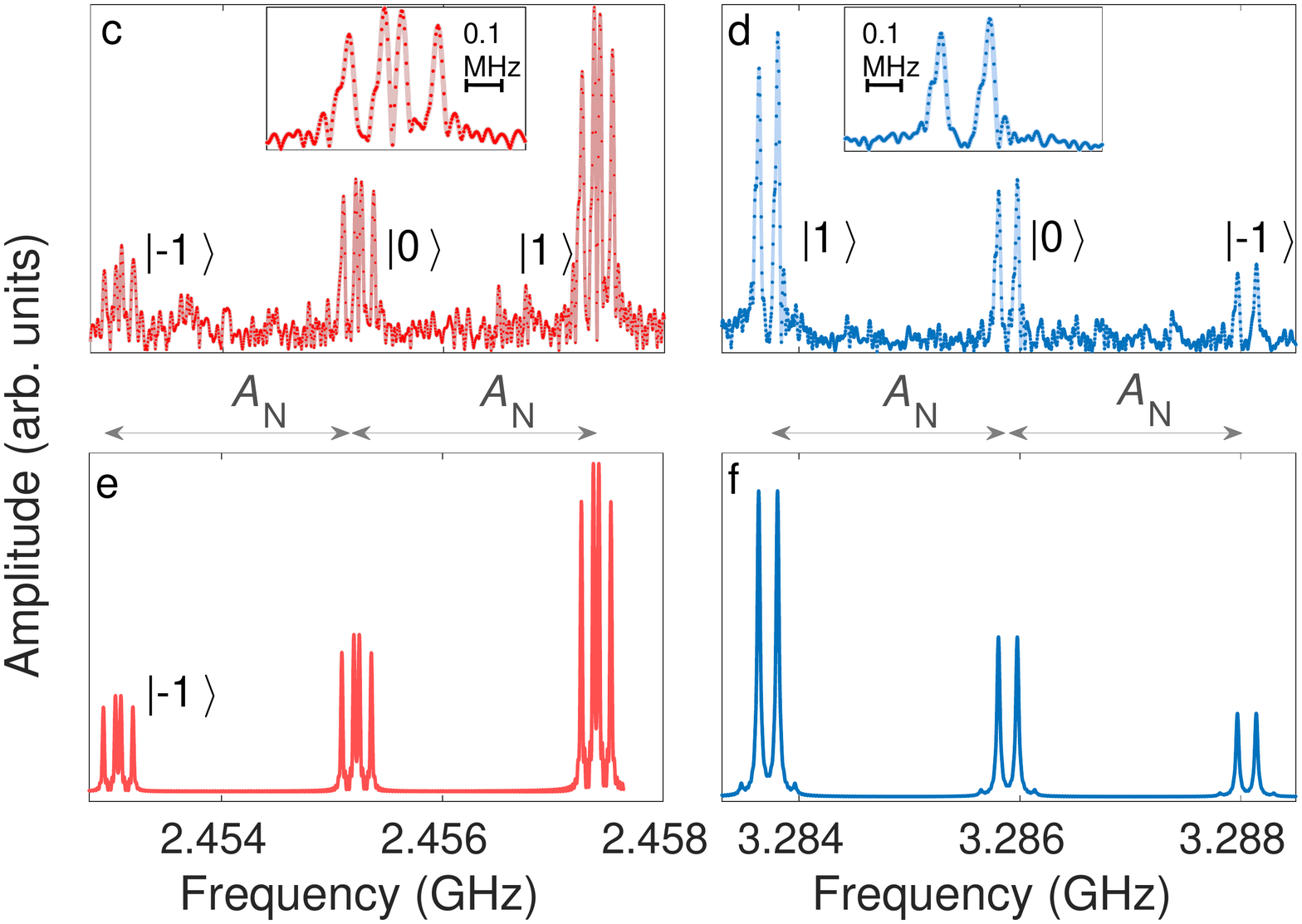} 
\caption{Characteristics of the system. (a) Structure of the NV system with
the electron spin coupled to one $^{14}$N and one $^{13}$C nuclear
spin. (b) Energy levels of the system consisting of the electron,
$^{14}$N and $^{13}$C nuclear spins. The vertical double arrows
indicate the ESR transitions that we use in this work. (c, d) Experimental
ESR spectra obtained in a field of $B=14.8$ mT for the transitions
between the $m_{S}=0\leftrightarrow-1$ and $m_{S}=0\leftrightarrow+1$
levels, respectively. The labels $|0\rangle$, $|1\rangle$, and $|-1\rangle$
mark the state of the $^{14}$N spin. The insets show the peaks corresponding
to the $^{14}$N state $|1\rangle$ on an expanded scale. The horizontal
double arrows indicate the hyperfine coupling with $^{14}$N. The
spectra in (e, f) show the matching numerical simulations. The small
outer peaks in the simulated spectrum in figure (f) are below the
noise levels in the experimental spectrum.}
\label{figreswhole} 
\end{figure*}

We consider the spin system consisting of the electron, the $^{14}$N
nuclear spin and one $^{13}$C nuclear spin. The axis of the NV system,
together with the position of the $^{13}$C nucleus define a symmetry
plane for the system \cite{PhysRevB.94.060101}. We therefore use
a symmetry-adapted coordinate system, where the $z$-axis is oriented
along the NV axis, while the $^{13}$C nucleus is located in the $xz$-plane,
as shown in Fig. \ref{figreswhole} (a). The static magnetic field
$\overrightarrow{B}$ is aligned along the NV symmetry axis so as
to simplify the Hamiltonian and to maximize the electron spin polarization
\cite{NJPmaxP}. The relevant Hamiltonian is then 
\begin{eqnarray}
\mathcal{H}/(2\pi) & = & DS_{z}^{2}-\nu_{e}S_{z}+P(I_{z}^{N})^{2}-\nu_{N}I_{z}^{N}-\nu_{C}I_{z}\nonumber \\
 & + & A_{N}S_{z}I_{z}^{N}+A_{zz}S_{z}I_{z}+A_{zx}S_{z}I_{x}.\label{eq:Ham3Spins}
\end{eqnarray}
The symbols for the electron spin $\vec{S}$ and the $^{13}$C nuclear
spin $\vec{I}$ are defined in the main text. $I_{z}^{N}$ denotes
the spin-1 operator for $^{14}$N, which experiences a nuclear quadrupole
splitting with coupling constant $P=-4.95$ MHz and a hyperfine interaction
with coupling constant $A_{N}=-2.16$ MHz ~\cite{PhysRevB.89.205202,PhysRevB.47.8816,Yavkin16}.
Fig. \ref{figreswhole} (b-f) show the energy levels and ESR spectra.

If we focus on a subspace where the state of the $^{14}$N is fixed
($m_{N}=1$ in the main text), $\mathcal{H}_{e,C}$ can be diagonalized
by the unitary transformation 
\begin{equation}
U_{T}=|1\rangle\langle1|\otimes R_{y}(\theta_{+})+|0\rangle\langle0|\otimes E+|-1\rangle\langle-1|\otimes R_{y}(\theta_{-}),\label{Ut3}
\end{equation}
where $R_{y}(\theta_{\pm})=e^{-i\theta_{\pm}I_{y}}$. The four ESR
transitions $m_{S}=0\leftrightarrow\pm1$ appear then $D\mp(\nu_{e}-A_{N})$,
shifted by 
\begin{equation}
\frac{1}{2}(\nu_{\pm}+\nu_{C}),\hspace{0.1cm}-\frac{1}{2}(\nu_{\pm}-\nu_{C}),\hspace{0.1cm}\frac{1}{2}(\nu_{\pm}-\nu_{C}),\hspace{0.1cm}-\frac{1}{2}(\nu_{\pm}+\nu_{C}),\label{tran113}
\end{equation}
where the upper / lower sign indicates that they are associated with
the $m_{S}=\pm1$ states, respectively. The corresponding transition
probabilities are 
\begin{equation}
\sin^{2}(\theta_{\pm}/2),\hspace{0.1cm}\cos^{2}(\theta_{\pm}/2),\hspace{0.1cm}\cos^{2}(\theta_{\pm}/2),\hspace{0.1cm}\sin^{2}(\theta_{\pm}/2)\label{prob11}
\end{equation}
respectively.

\subsection{Pulse Sequences}

\begin{figure}[b]
\includegraphics[width=1\columnwidth]{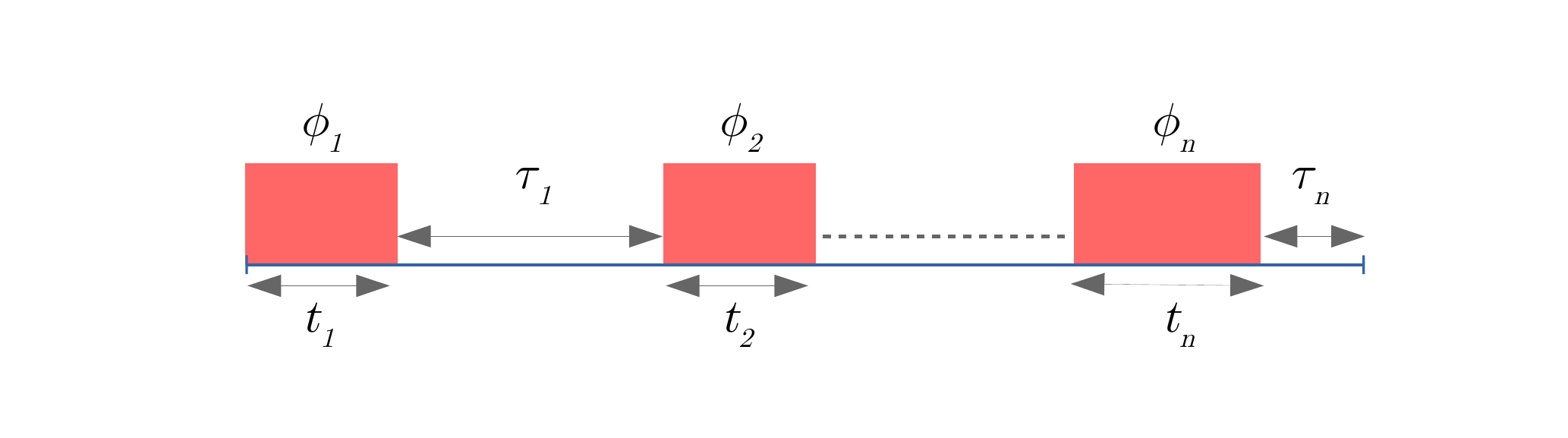} \caption{Pulse sequence and variables considered for the numerical search.
Here $t_{1},\cdots,t_{n}$ are the durations and $\phi_{1},\cdots,\phi_{n}$
the phases of the MW pulses and $\tau_{1},\cdots,\tau_{n}$ are the
delays between them. The amplitude of the pulses is fixed with a Rabi
frequency $\omega_{1}/(2\pi)$ = 0.5 MHz.}
\label{opt_pulse} 
\end{figure}

In a subspace spanned by the states 
\begin{equation}
\{|0\rangle,|-1\rangle\}_{e}\otimes\{|1\rangle\}_{N}\otimes\{|\uparrow\rangle,|\downarrow\rangle\}_{C},
\end{equation}
\label{basissub} the Hamiltonian of the electron-$^{13}$C system
can be represented as 
\begin{equation}
\frac{1}{2\pi}\mathcal{H}_{s}=(-\nu_{C}-\frac{A_{zz}}{2})I_{z}+A_{zz}s_{z}I_{z}+A_{zx}s_{z}I_{x}-\frac{A_{zx}}{2}I_{x}\label{Hsub}
\end{equation}
in the rotating frame with frequency $D+\nu_{e}-A_{N}$, where $s_{z}$
denotes the pseudo-spin 1/2 operator for electron spin. We consider
pulse sequences consisting of $n$ MW pulses with fixed Rabi frequency
$\omega_{1}$, as shown in Fig. \ref{opt_pulse}. The propagators
for the individual MW pulses can be written as $U_{k}^{MW}=e^{-i\mathcal{H}_{k}^{MW}t_{k}}$
where $\mathcal{H}_{k}^{MW}=\mathcal{H}_{s}+\omega_{1}[s_{x}\cos(\phi_{k})+s_{y}\sin(\phi_{k})]$,
and for the free evolutions as $U_{k}^{d}=e^{-i\mathcal{H}_{s}\tau_{k}}$,
with $k=\{1,\cdots,n\}$. The total unitary $U$ is a time ordered
product of the $U_{k}^{MW}$ and $U_{k}^{d}$, and is a function of
the pulse parameters $(t_{1},\cdots,t_{n},\phi_{1},\cdots,\phi_{n},\tau_{1},\cdots,\tau_{n})$.
The goal is to design $U$ with suitable pulse parameters such that
the fidelity $F_{g}=|\mathrm{Tr(U_{T}^{\dagger}U)}|/4$ of the effective
propagator $U$ with respect to the target propagator $U_{T}$ is
maximized. We used a genetic algorithm as a numerical search method
to obtain the best pulse parameters.

For some applications, we do not have to find a specific unitary propagator,
but it is sufficient to transfer a given initial state $\rho_{0}$
to a target state $\rho_{T}$. The actual final state is then $\rho_{opt}=U\rho_{0}U^{\dagger}$,
where $\rho_{opt}\equiv\rho_{opt}(t_{1},\cdots,t_{n},\phi_{1},\cdots,\phi_{n},\tau_{1},\cdots,\tau_{n})$,
and we maximize the state fidelity $F_{s}=\mathrm{Tr}(\rho_{T}\rho_{opt})/\sqrt{\mathrm{Tr}(\rho_{T}^{2})\mathrm{Tr}(\rho_{opt}^{2})}$.

\begin{figure}
\includegraphics[width=1\columnwidth]{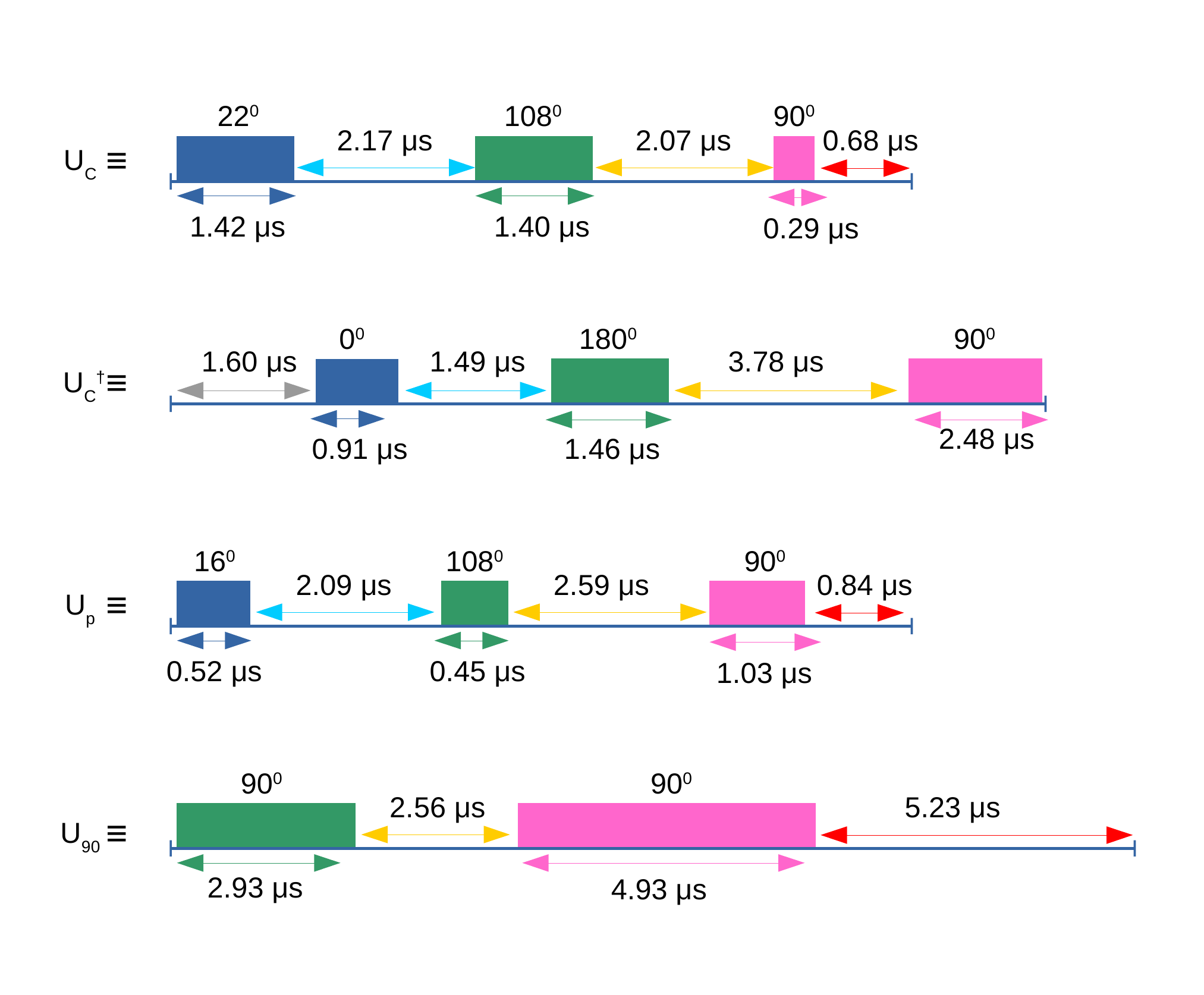} \caption{Pulse sequences for $U_{c}$, $U_{c}^{\dagger}$, $U_{p}$ and $U_{90}$
at a fixed $\omega_{1}/(2\pi)$ = $0.5$ MHz. The MW pulses are solid
rectangles and are resonant with the ESR transition $0\leftrightarrow-1$.
The phases of the pulses are indicated on the top of the rectangles.}
\label{Uc_pulse} 
\end{figure}

The performance of the pulses is sensitive to variations in $\omega_{1}$.
To obtain good fidelity in experiments where the actual MW amplitude
deviates from the ideal one, we optimized the pulse sequences for
a range of amplitudes, taking the average fidelity as the performance
measure. For the gates $U_{c}$, $U_{c}^{\dagger}$ and $U_{p}$,
we used the range $\omega_{1}/2\pi=[0.47,0.53]$ MHz and for $U_{90}$
$\omega_{1}/2\pi=[0.48,0.52]$ MHz. The optimized pulse sequences
for $U_{c}$, $U_{c}^{\dagger}$, $U_{p}$ and $U_{90}$ are shown
in Fig. \ref{Uc_pulse}. The theoretical robust state fidelities for
the four sequences are $95\%$, $95\%$, $98\%$ and $92\%$, respectively.

\begin{figure}
\includegraphics[width=1\columnwidth]{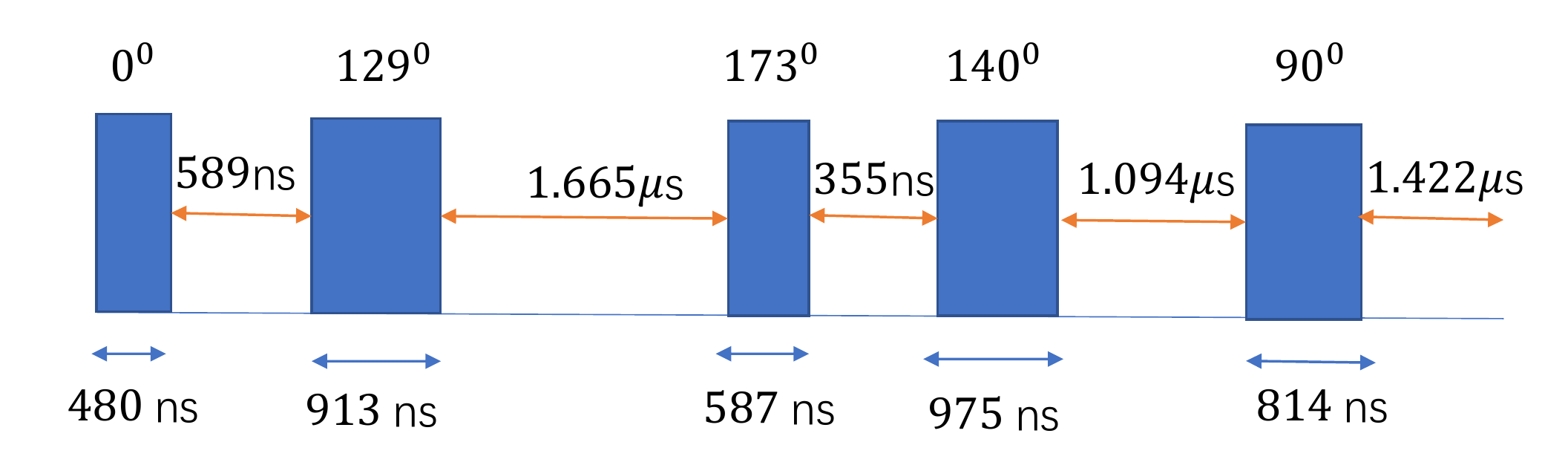} \caption{Pulse sequence for $U_{p}$ for $\theta_{-}=36.6^{\circ}$, when $\nu_{C}=0.3$
MHz, at a fixed $\omega_{1}/(2\pi)$ = $0.5$ MHz. }
\label{Figsmall}
\end{figure}

In Fig. \ref{Figsmall}, we illustrate a pulse sequence to implement
$U_{p}$ for the case of $\nu_{C}=0.3$ MHz, by increasing the static
magnetic field.

\subsection{Experimental performance of $U_{90}$ and $U_{c}$}

\subsubsection{$U_{90}$}

\begin{figure}
\centering %
\begin{tabular}{c}
\includegraphics[width=0.8\columnwidth]{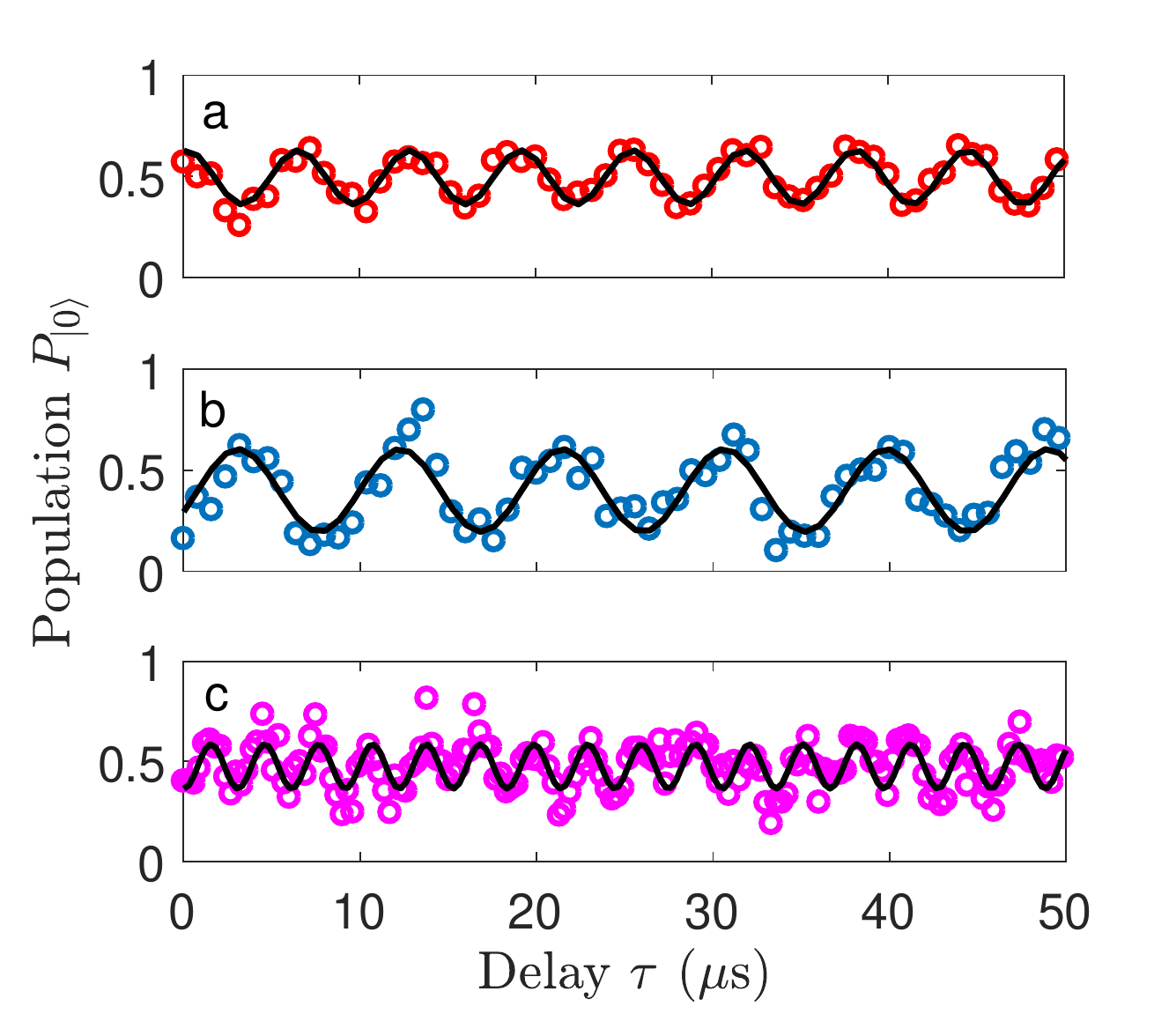} \tabularnewline
\end{tabular}\caption{The initial parts of the FIDs shown in Fig. \ref{fid_exp}, with (a-c)
corresponding to Figs. \ref{fid_exp} (a), (c) and (e), respectively. }
\label{fid_expa3}
\end{figure}

We can exploit the measured populations $P_{|0\rangle}$ shown in
Figs. \ref{fid_exp} (a), (c) and (e) to estimate experimental fidelity
of $U_{90}$. Here we chose the data in initials parts, since the
signals in these periods are less noisy. Then we fit the data points
using a function as $a_{m_{S}}+b_{m_{S}}\sin(2\pi\nu_{m_{S}}\tau+c_{m_{S}})$,
shown in Fig. \ref{fid_expa3}. Here $\nu_{m_{S}}$ denotes the transition
frequency measured in the spectra shown in Figs. \ref{fid_exp}(b,
d, f), respectively, with the index indicating the involved subspace.
The parameters $a_{m_{S}}$, $b_{m_{S}}$ and $c_{m_{S}}$ are constants
in each subspace.

We used the ratios between $b_{m_{S}}$ to estimate the experimental
fidelities for $U_{90}$ as well as the $180^{\circ}$ MW pulse in
the insets in Figs. \ref{fid_exp}(d-f) as the following. In this
manner, we can eliminate the errors in the operation of polarizing
the $^{13}$C spin and the detection operation $U_{t}$. The fitted
$b_{m_{S}}$ is listed as $0.13$, $0.11$ and $0.20$ for $b_{m_{S}}=0$,
$1$ and $-1$, respectively. Noticing the pulse sequences shown as
the insets in Figs. \ref{fid_exp}(b) and (f), we can obtain the fidelity
for the $180^{\circ}$ MW pulse in the inset (f) as $F_{180}=\sqrt{b_{1}/b_{0}}=0.92$.
In a similar way, we then obtained the fidelity for $U_{90}$ as $F_{U_{90}}=\sqrt{b_{1}/b_{-1}}=0.74$.
Here we assumed that the the $180^{\circ}$ pulses in insets (d-f)
have the same fidelity.

\subsubsection{$U_{c}$}

\begin{figure}
\centering %
\begin{tabular}{c}
\includegraphics[width=1\columnwidth]{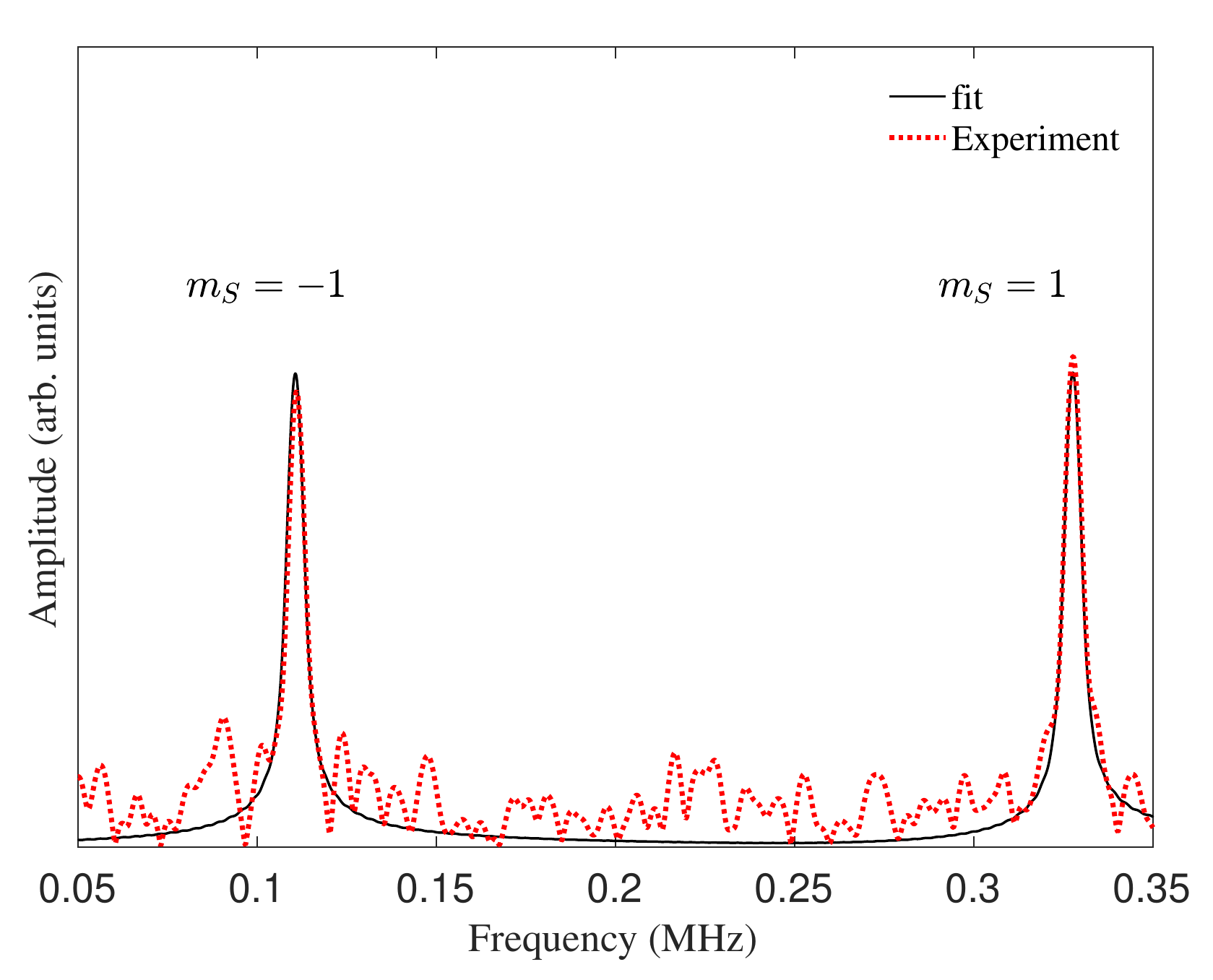} \tabularnewline
\end{tabular}\caption{Experimental and theoretical $^{13}$C spin spectra corresponding
to the pulse sequence in Fig. \ref{figfidsingle}(f) of the main text.
Here the theoretical spectrum is scaled by a factor of 0.7 to fit
the experimental spectra.}
\label{fid_expa1} 
\end{figure}

In this section, we evaluate the performance of the operation $U_{c}$
using the experimental $^{13}$C spin spectrum in Fig. \ref{figfidsingle}
(f) measured with the pulse sequence shown in Fig. \ref{figfidsingle}
(c). We then compare this experimental spectrum with the corresponding
theoretical spectrum. The theoretical population at the end of this
pulse sequence is $P_{|0\rangle}^{'}$ and its dependence on the delay
$\tau$ is explained as Eq. (\ref{eq:popUcp}) in the main manuscript.
The Fourier transformation of $P_{|0\rangle}^{'}$ in the frequency
domain is $y(\nu)=\mathcal{F}\{P_{|0\rangle}^{'}\}$. We fit this
theoretical spectrum to the experimental spectrum by multiplying $y(\nu)$
by a factor $f=0.7$. This fit is shown in Fig. \ref{fid_expa1}.
Following this result, we can estimate the experimental fidelity for
$U_{c}$ and $U_{c}^{\dagger}$ as $F_{U_{c}}=\sqrt{f}/F_{180}=0.91$,
where $U_{c}^{\dagger}$ is assumed the same as $U_{c}$.

\subsection{Other supplementary data}

In Table \ref{blochxyz}, we list the components of the final states
on the Bloch sphere shown in Fig. 1.

\begin{figure}
\centering %
\begin{tabular}{c}
\includegraphics[width=1\columnwidth]{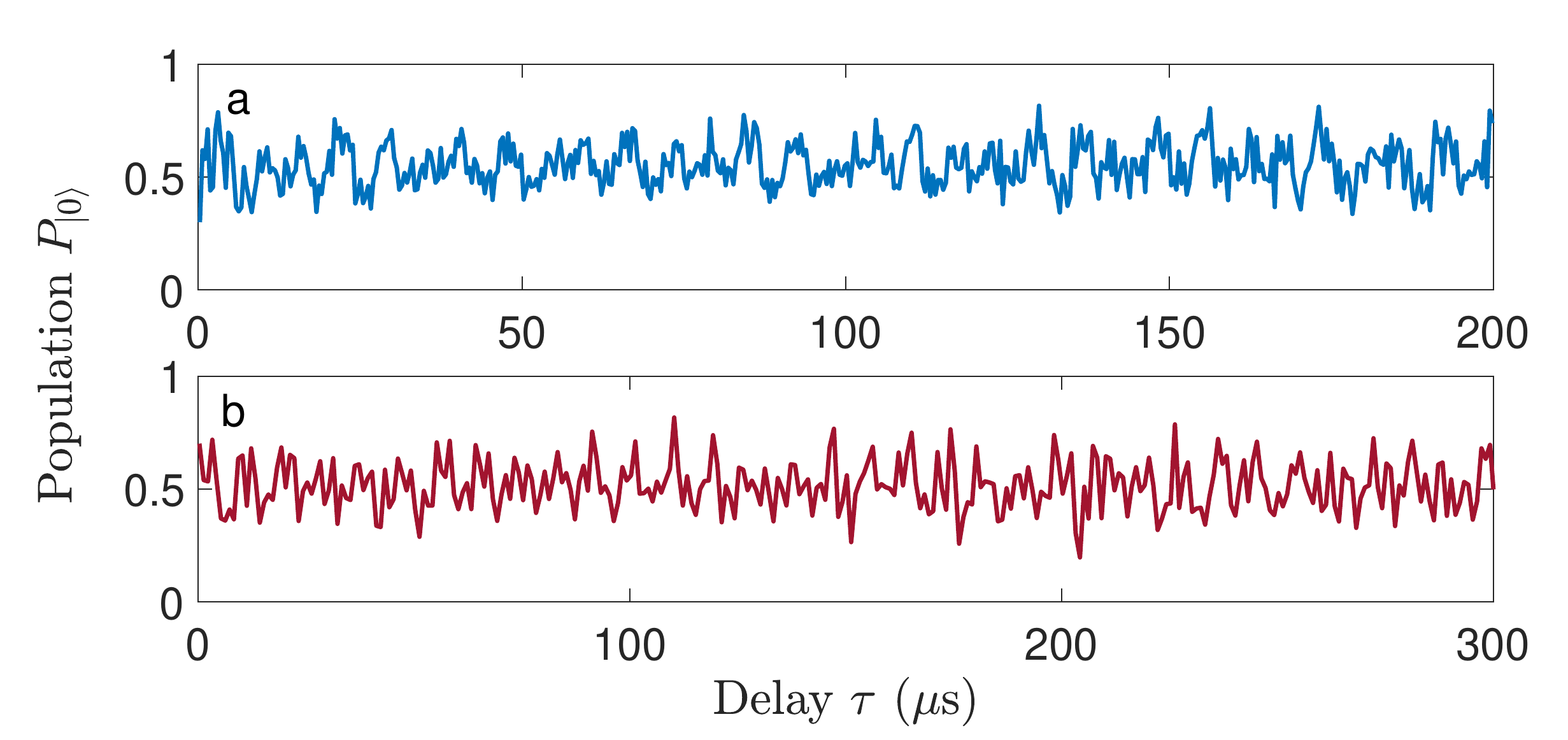} \tabularnewline
\end{tabular}\caption{(a-b) $^{13}$C FID signals obtained by the pulse sequences shown
in Figs. \ref{figfidsingle} (a) and (c). The spectra shown in Fig.
\ref{figfidsingle} (e-f) were obtained by Fourier transformation
of these signals. }
\label{fid_expa4} 
\end{figure}

In Figs. \ref{fid_expa4} (a-b), we illustrated the FID signals of
$^{13}$C obtained by the pulse sequences shown as Figs. \ref{figfidsingle}
(a) and (c), respectively. The corresponding spectra are shown in
Fig. \ref{figfidsingle} (e-f). In experiment we optimized the duration
of the FID signal as 200 or 300 $\mu$ for higher S/N. Since the duration
of the FID signals is much shorter than both the longitudinal relaxation
time of the electron spin ($T_{1}^{e}\approx3.5$ ms \cite{Ournewpaper}),
and traversal relaxation time of $^{13}$C (estimated longer than
$T_{1}^{e}$ \cite{PhysRevA.87.012301}), we cannot observe the clear
decay of the signals. In the process of Fourier transformation, we
added certain window functions and zero filling to improve the spectra.
Therefore we cannot observe such relaxation times from the linewidths
of the resonance lines in Figs. \ref{figfidsingle} (e-f).

\begin{table}
\begin{tabular}{|c|c|c|c|c|}
\hline 
Operation  & Spin  & $x$  & $y$  & $z$ \tabularnewline
\hline 
$U_{c}$  & Electron  & -0.0068  & 0.1272  & 0.0008 \tabularnewline
\hline 
$U_{c}$  & $^{13}$C  & -0.0788  & -0.3626  & 0.4734 \tabularnewline
\hline 
$U_{p}$  & Electron  & 0.0004  & -0.0016  & 0.0004 \tabularnewline
\hline 
$U_{p}$  & $^{13}$C  & 0.0698  & -0.0608  & 0.9920 \tabularnewline
\hline 
$U_{90}$  & Electron  & 0.0160  & -0.0622  & 0.9690 \tabularnewline
\hline 
$U_{90}$  & $^{13}$C  & 0.1618  & -0.9154  & 0.2810 \tabularnewline
\hline 
\end{tabular}\caption{The $x$, $y$ and $z$ components of the final states on the Bloch
sphere shown in Fig. 1.}
\label{blochxyz} 
\end{table}

\end{document}